\documentclass[pra, preprint, floatfix, amsmath, superscriptaddress, nofootinbib]{revtex4-1}

\usepackage{graphicx, color}
\usepackage{amsmath}
\usepackage{amssymb}
\usepackage{amsthm}
\usepackage{amsfonts}
\usepackage[english]{babel}
\usepackage{url}
\usepackage{multirow}
\usepackage{ulem}
\usepackage{float}

\begin{document}
\title{Deterministic and controllable photonic scattering media via direct laser writing}
\author{E. Marakis}
\affiliation{Complex Photonic Systems (COPS), MESA+ Institute for Nanotechnology, University of Twente, PO Box 217, 7500 AE Enschede, The Netherlands}
\author{ R. Uppu}
\altaffiliation{Present address: Niels Bohr Institute, University of Copenhagen, Blegdamsvej 17, 2100 Copenhagen, Denmark}
\author{M. L. Meretska}
\altaffiliation{Present address: Harvard John A. Paulson School of Engineering and Applied Sciences, Harvard University, Cambridge, Massachusetts 02138, United States.}
\author{K. J. Gorter, W. L. Vos and P. W. H. Pinkse}
\affiliation{Complex Photonic Systems (COPS), MESA+ Institute for Nanotechnology, University of Twente, PO Box 217, 7500 AE Enschede, The Netherlands}

\nocite{bibid}
\begin{abstract}
Photonic scattering materials, such as biological tissue and white paper, are made of randomly positioned nanoscale inhomogeneities in refractive index that lead to multiple scattering of light.
Typically these materials, both naturally-occurring or man-made, are formed through self assembly of the scattering inhomogeneities, making it extremely challenging to know the exact positions of these inhomogeneities, let alone control those.
Here, we report on the nanofabrication of photonic multiple-scattering media using direct laser writing with a deterministic design.
These deterministic multiple-scattering media consist of submicron thick polymer nanorods that are randomly oriented within a cubic volume.
We study the total transmission of light as a function of the number of rods and of the sample thickness to extract the scattering and transport mean free paths using radiative transfer theory.
Such ability to fabricate photonic multiple-scattering media with deterministic and controllable properties opens up a myriad of opportunities for fundamental studies of light scattering, in particular in the multiple-scattering regime and with strong anisotropy and for new applications in solid-state lighting and photovoltaics.
\end{abstract}
\maketitle
\section{Introduction}
%
%
The scattering of light is a familiar physical process that is abundant in everyday life and in nature; manifestations of scattering are the opacity of white paper, of clouds, and of biological tissue~\cite{Goodman1976, Ishimaru1978,vanAlbada1985,Wolf1985, Genack1987,Cheong1990,Garcia1992,Mishchenko1993, Lagendijk1996,vanRossum1999,Akkermans2007, Wiersma2013,Vos2013,Burresi2013,Rotter2017,Meretska2017}.
Light scattering occurs at any interface between different materials and causes a part of the light to deviate from its original path. 
A sufficiently thick medium with a high density of such interfaces acts as a multiple-scattering material.
The light transport through and interaction with the scattering medium is characterized by two characteristic length scales: the scattering mean free path $\ell\textsubscript{sc}$ and the transport mean free path $\ell\textsubscript{tr}$.
The scattering mean free path quantifies the mean distance between subsequent scattering events. 
The mean distance that the incident light propagates before its direction is scrambled is the transport mean free path $\ell\textsubscript{tr}$. 
 Scattering media with $L > \ell_\textrm{tr} > \lambda$, where $L$ is its physical thickness and $\lambda$ is the wavelength of light in the embedding medium, support diffusive transport of light and play an important role in devices such as solar cells and white LEDs.
For instance, a thin scattering layer on top of a solar cell increases the absorption of sunlight and hence the cell's efficiency \cite{Burresi2013}.
White LEDs employ a layer of scattering and phosphor particles to control the spectral and spatial distribution of the emission \cite{Vos2013, Meretska2017}.
The aforementioned applications of scattering media often require a specific optical thickness, defined as the ratio
$L/\ell\textsubscript{tr}$, and a carefully designed angular distribution of light quantified through the scattering anisotropy $g$ \cite{Jacucci2019AOM}. 
For example, high opacity diffusing media that could benefit white LEDs and thin film solar cells require that the physical thickness $L$ exceeds the scattering mean free path ($L/\ell\textsubscript{sc} >1$) for enhanced absorption, while ensuring directional illumination or minimal backscattering, i.e. $\ell\textsubscript{tr} \approx L$.
Importantly, such scattering media benefit from physical thickness down to a few $\mu$m, necessitating
short $\ell\textsubscript{sc}\approx\mu$m. 
In a non-absorbing scattering medium, $g$ relates the mean free paths as $(1-g)\ell\textsubscript{tr}=\ell\textsubscript{sc}$, hence the LEDs and solar cells require designer scattering media with short $\ell\textsubscript{sc}$ and large $g$.


A large number of scattering media with well-defined scattering properties, covering a broad range of optical thicknesses have been presented in the literature \cite{Garcia1989,Cao1998,Schuurmans1999science,Schuurmans1999prl,Garcia2010,Miranda2017AOM}. However, the fabrication of those media rely on random processes and only the average values are controlled during fabrication. 
Knowledge about the internal structure or actually tailoring the internal structure on the microscopic scale would give an entirely new level of control of the scattering  properties, e.g. by employing the rapidly evolving inverse design techniques \cite{Molesky2018natphot,Christiansen2019prl}. 
However, revealing the microscopic internal structure of multiple-scattering media requires careful dissection of the media for sub-micron inspection, thereby destroying the medium in the process.
%
Recent advances in 3D nanofabrication \cite{Farsari2010,Fischer2013,Seniutinas2018} has made it feasible to create a scattering medium with a precisely known predetermined internal structure. 
A priori knowledge of the internal structure makes it ideal for studies of light propagation that are sensitive to microscopic wave interference of light \cite{Utel2019}.
In the literature, several scattering media with deterministic geometry and scattering properties have been demonstrated using different nanofabrication techniques. 
Matoba et al.\,\cite{Matoba2009} have created a large-volume deterministic scattering medium with laser micromachining \cite{Gattass2008,Battista2016}, but with undefined scattering strength. 
An alternative nanofabrication technique is direct laser writing (DLW) \cite{Fischer2013}, offering about $1$--$10\,$nm material deposition precision and a feature size down to about 100\,nm \cite{Farsari2010}. 
DLW encompasses many technical variations and definitions \cite{Thiel2010,Do2013,Kiefer2020}. In the present work we focus on DLW employing multiphoton lithography. The physical principle of this type of DLW is multiphoton
absorption, which initiates polymerization at a targeted location inside the volume of the photoresist. The advantage of DLW is the freedom to create deterministic and complex 3D geometries.
An example of a DLW-nanofabricated photonic structure with complex geometry is the hyperuniform %
medium 
fabricated by Haberko et al. \cite{Haberko2013pra,Haberko2013opex,Muller2014}. 
In the current work, we report on the fabrication of deterministic scattering media that combine both a large anisotropy and a short (few micrometer)  scattering mean free path. We measure the transport mean free path of the structures and validate that light scattering in our structures is in the diffusive regime ($L/\ell\textsubscript{tr} >1$).
%
%
%
\section{Designing a deterministic scattering medium}

\begin{figure}[tbp!]
\includegraphics[width=0.75\linewidth]{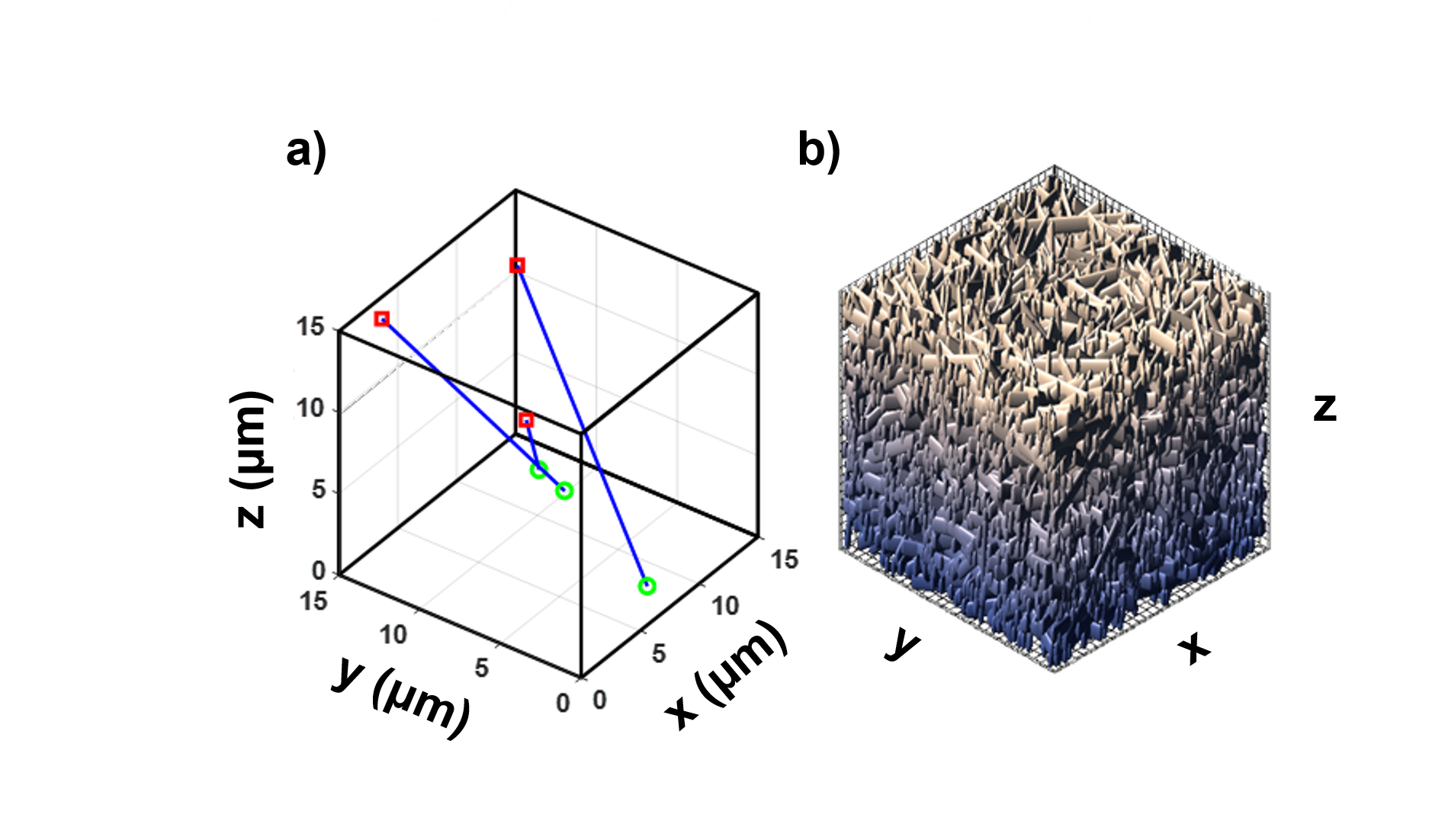}
\caption{The three-dimensional spatial structure of our deterministic scattering samples. 
a) First rods are connected to the substrate (green points) in order to make a rigid connected scaffold. 
The end points lie on another face of the cube (red points).
Secondly, rods that are not connected to the substrate are written subsequently. 
This procedure minimizes drifting of the features during the writing process. 
b) A rendering of the final model. 
The features on top have a lighter shading, while the features closer to the substrate are darker.}
\label{fig:proc}
\end{figure}
We realize direct laser-written disordered scattering media as a collection of randomly-oriented polymer rods in a cubic volume.
As described in Ref.~\cite{Marakis2019}, we developed an algorithm based on Jaynes' solution to Bertrand's paradox to ensure an on-average uniform filling of the volume and the random positioning of the rods in the cube with the use of a random number generator.
A design example is shown in \textbf{Figure\,\ref{fig:proc}}(b). 
The randomly intersecting polymer rods run from one facet of the cube to another.
Typically, creating mechanically stable structures using DLW requires additional support walls bounding the structure.
However, these walls create unwanted light scattering interfaces that complicate the modeling of light transport.
Our novel design achieves mechanical stability of the structures without the addition of any support walls, and hence enables direct comparison with theoretical models of light scattering that are necessary for extracting the transport parameters ($\ell\textsubscript{sc}, \ell\textsubscript{tr}, g$).
The DLW setup is ideally suited to write rods in a successive manner \cite{Sun2004}. 
After generating the coordinates of the rods, the algorithm sorts them such that the rods attached to the substrate are written first, as illustrated in Figure\,\ref{fig:proc}(a). 
This creates a stable scaffold to which the other rods are attached to. 
At the end of the writing process, a rigid structure is created to follow the design model as shown in Figure\,\ref{fig:proc}(b).

\begin{figure}[tbp!]
\begin{center}
\includegraphics[width=0.5\linewidth]{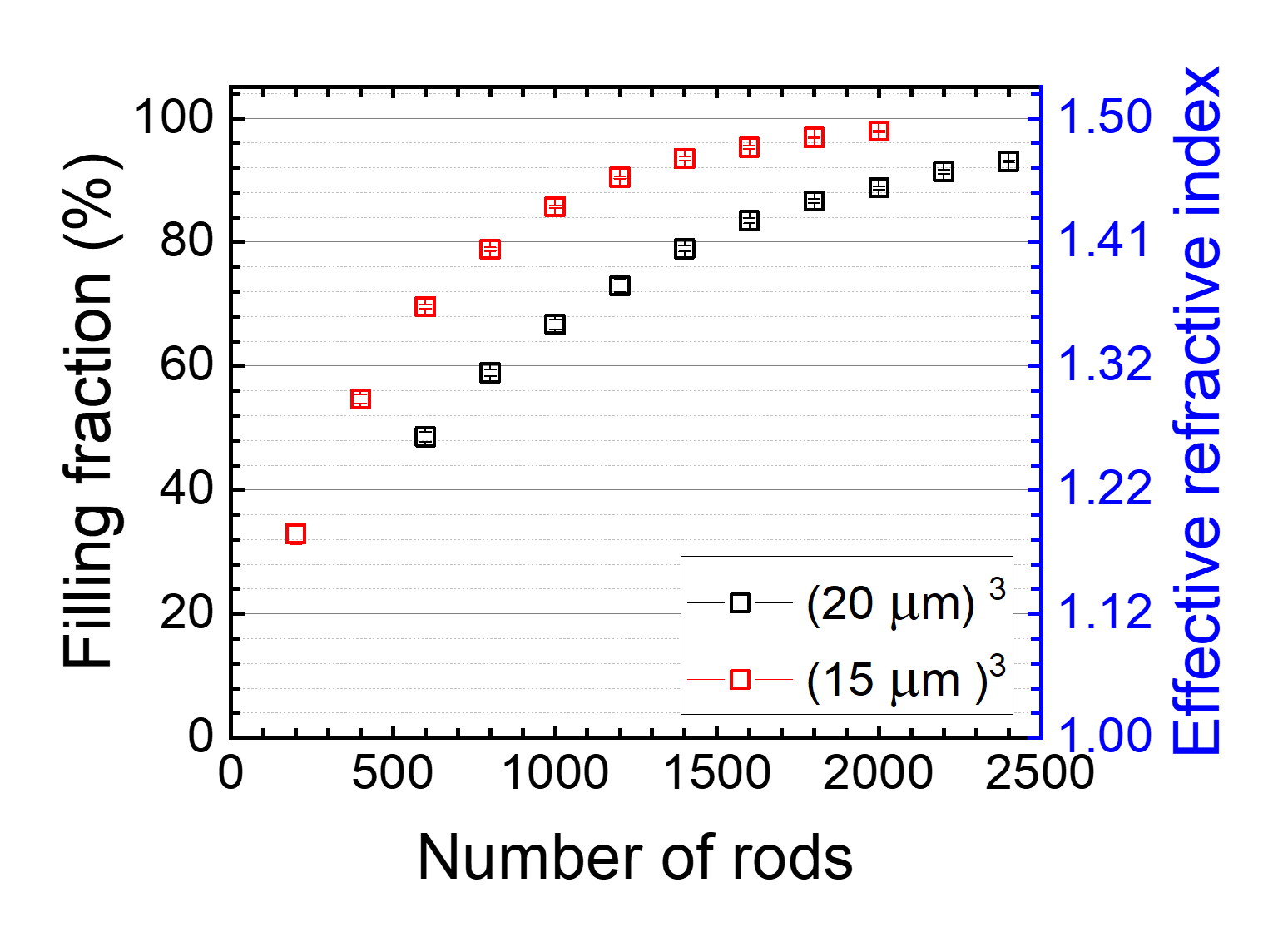}
\caption{Estimated filling fraction and effective refractive index versus the density of rods over volume.
Red markers and line denote how the filling fraction increases with increasing density of rods per volume for structures with volume (15\,$\mu m$)\textsuperscript{3}, while the black markers show the same trend for cubes of (20\,$\mu m$)\textsuperscript{3}. 
The estimation was performed by numerically integrating the volume of the intersecting rods.}
\label{fig:fill_neff}
\end{center}
\end{figure}

An important parameter {\color{black} that determines the scattering strength} of the disordered medium is the filling fraction of polymer, which is set by the number of rods written in the cubic volume. 
Starting from an empty cube, the scattering strength will increase with the number of rods until at some density of rods, the overlapping of the different rods will effectively create a solid block of polymer with a decreasing volume of voids, decreasing the scattering strength of the structure. 
The desired filling fraction is the one that maximizes light scattering. 
Creating a free-standing structure requires a minimum number of rods that intersect to form a rigid skeleton that does not collapse under its own weight or under capillary forces. 
When the number of rods are increased above a certain critical value, the air-polymer surface will decrease, thereby reducing the scattering strength of our structures.
\textbf{Figure\,\ref{fig:fill_neff}} shows the estimated filling fraction and effective refractive index\cite{Vos2001} of the structures with increasing number of rods for two fixed structure volumes. 
The deterministic fabrication of our structures allows us to calculate the filling fraction of the structures, given the knowledge of their design \cite{footnote}. 
The estimated filling fraction also allows us to estimate the effective refractive index for a wavelength of 633\,nm, indicated in Figure\,\ref{fig:fill_neff} by the second ordinate. Hereto we used a simple volume average of the dielectric constants.
A higher number of rods is required to reach the same filling fraction in the (20\,$\mu$m)\textsuperscript{3} structure in comparison to the (15\,$\mu$m)\textsuperscript{3}, which highlights isotropic disorder in the structure.

To keep fabrication times acceptable (7 minutes per structure), we created structures with a lateral size of 15 to 20\,$\mu $m and a height of up to 20\,$\mu $m. 
For structures wider and taller than 20\,$\mu $m, we suggest the model to be divided into cubes of (20\,$\mu $m)\textsuperscript{3} to prevent pyramidal distortions, shadowing effects, and weight deformations \cite{Zhou2015,Bauhofer2017}.

\section{Results}

\subsection{Fabrication quality}

\begin{figure}[tbp!]
\begin{center}
\includegraphics[width=0.5\textwidth]{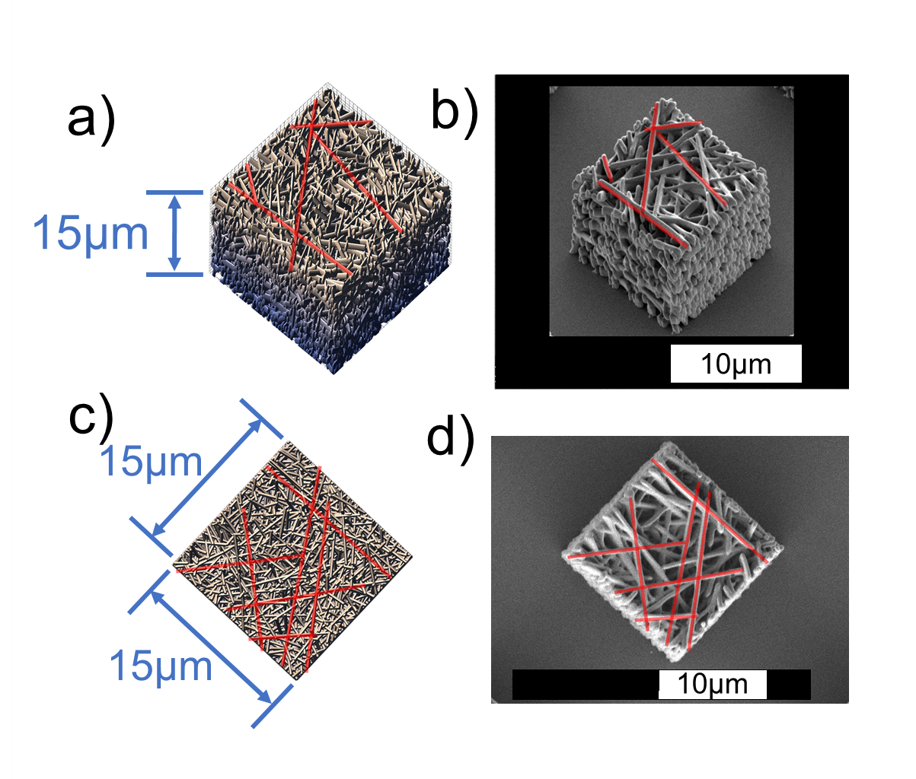}
\caption{Design of a scattering sample of polymer rods a) in bird's eye view and c) from above. b) and d) show SEM images of the fabricated structures according to the designs a) and c). We have highlighted several nanorods to ease the comparison of the model and the fabricated sample. }
\label{fig:sem}
\end{center}
\end{figure}

\begin{figure}[tbp!]
\begin{center}
\includegraphics[width=0.4\textwidth]{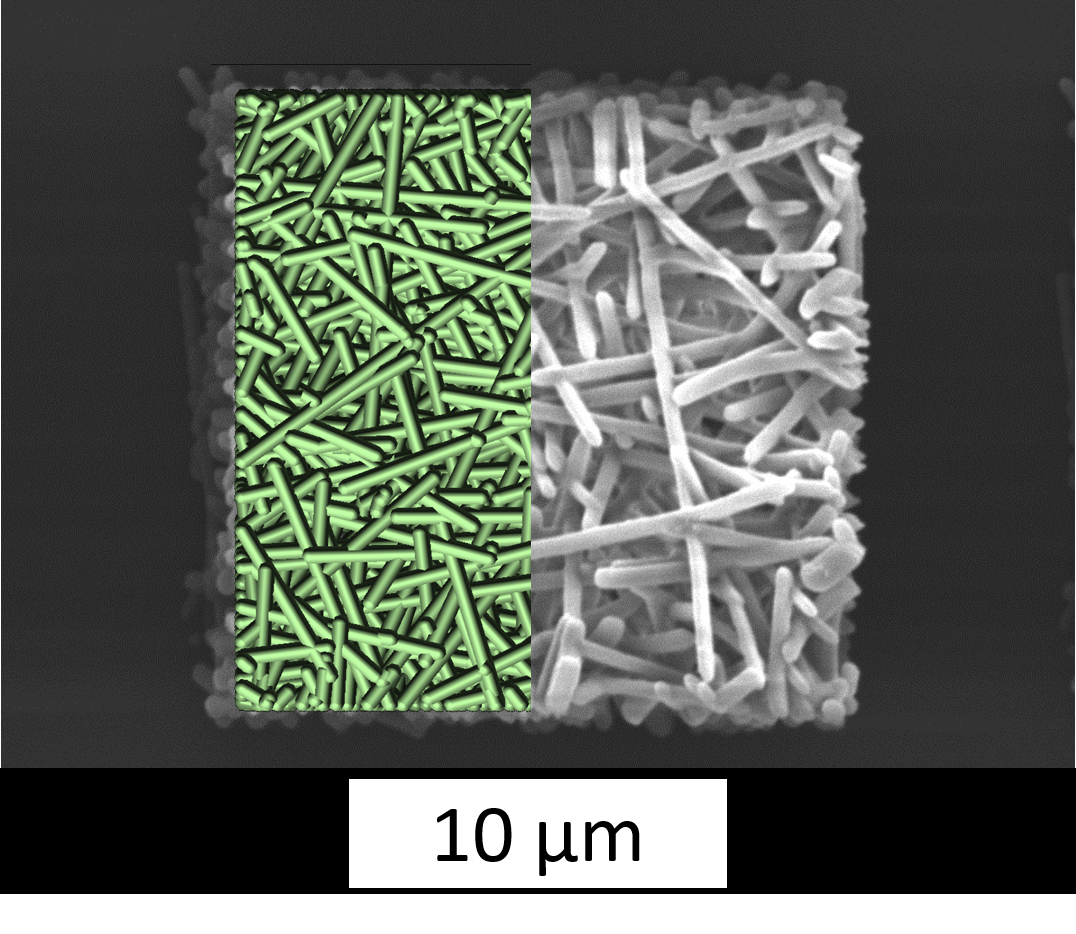}
		\caption{Top view of model and SEM picture of the fabricated structure. On can see that the features evolve continuously between model (shaded in green) and the fabrication (Grey SEM image). Some of the smaller rods may not adhere to the main skeleton, but their final contribution is small ($\sim$2\,vol\%). }
		\label{fig:comp_top}
	\end{center}
\end{figure}

We fabricated 693 structures with various thicknesses and filling fractions and employed a scanning electron microscope (SEM) to determine the feature sizes.
\textbf{Figure\,\ref{fig:sem}} shows a qualitative comparison of the surface features in the designed (a, c) and the fabricated structures (b, d) that highlights the similarity. Note that in contrast to the model rendering in (a) and (c), deeper lying lines in the SEM pictures (b) and (d) are not visible.
The rod thickness was measured to be 524 $\pm$ 60\,nm. The elongated focus of the laser beam in the resist results in an elliptical cross-section of the rods with the semi-major axis along $z$. The resulting size asymmetry of the voxel is 3.5.
We note that the elongation of a rod (i.e. its cross-sectional area) depends on its angle with respect to the z-axis \cite{Fischer2013}.
Although the resemblance between the design and real structure is very good and features remain in the same relative positions as seen in \textbf{Figure\,\ref{fig:comp_top}}, there are a few errors in the fabricated structure, for e.g. missing rods and shrinkage artifacts.
Typically, the very short rods on the cube faces wash away during development due to insufficient adhesion of these features to the structure.
The total number of missing rods is small (few dozen) and their contribution to the structure is very small, estimated to be less than 2\,vol\% and not easy to discriminate from the final structure as shown in Figure~\ref{fig:comp_top}.
Importantly, these fabrication errors repeat in different realization of the structure, thereby making it feasible to compare their optical interference properties.
We did not observe severe pyramidal distortions \cite{Haberko2013pra} from the inspection of the SEM images, a distortion that is common in direct laser-written structures \cite{Sun2004,Bauhofer2017}. 
 Although the residual pyramidal artifacts do decrease the density uniformity,  given the modest height of the structures, we expect this effect to have only a small impact on the optical measurements.
We conclude that the chosen height and laser dose were a good choice and proceed to the optical characterization of the structures.

\subsection{Scattering characteristics}

In order to study the total transmission and extract {\color{black}$\ell\textsubscript{tr}$,
}
we design a set of structures with varying height at a given filling fraction \cite{Garcia1992,VanDerMark1988}. 
The uniform density of the scatterers certifies an on-average uniform filling fraction also in the structures with a reduced height. 
{\color{black}
The microscopic design of the structures was kept constant and the 
height of the structures was tuned by writing deeper inside the glass substrate. 
This avoids the transmission noise that could be introduced by different disorder realizations.} 
In order to study the scattering properties of this design, we fabricated two series of structures with lateral dimensions of 15 and 20\,$\mu$m. 
The first series are made of $200$ up to $2000$ rods$/(15\,\mu$m$)^3$ with a step size of 200 rods$/(15\,\mu$m$)^3$, corresponding to a filling fraction from 32\,vol\% up to 97\,vol\%. 
For each filling fraction we also varied the height from 1.5\,$\mu$m up to 15\,$\mu$m with steps of 1.5\,$\mu$m. 
The second series is composed of 600 rods$/(20\,\mu $m$)^3$  up to 2400 rods$/(20\,\mu $m$)^3$  with a step size of 200 rods$/(20\,\mu $m$)^3$, corresponding to a filling fraction from 49\,vol\% up to 93\,vol\%. 
For each filling fraction we also varied the height from 6.5\,$\mu $m up to 20\,$\mu $m with a step of 1.5\,$\mu $m. 

\begin{figure}[tbp!]
	\begin{center}
\includegraphics[width=0.9\linewidth]{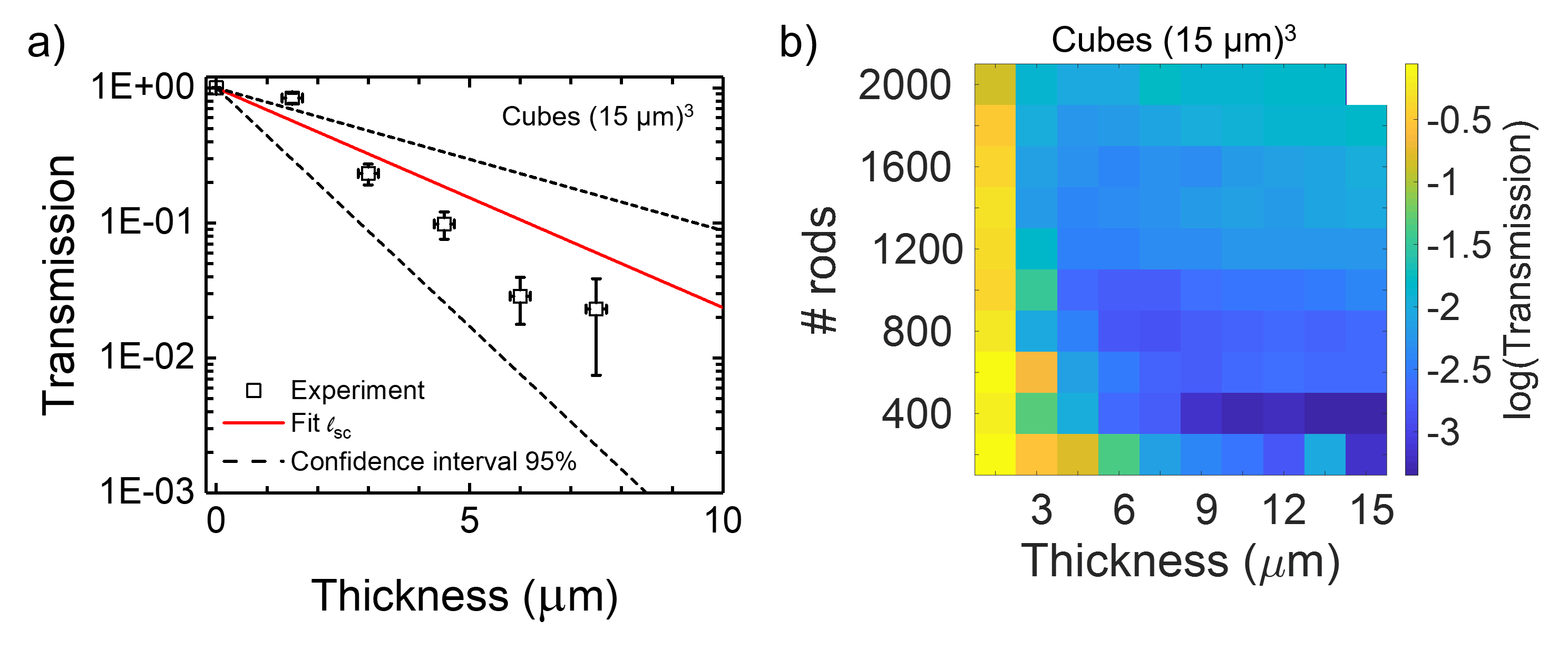}
\caption{ Ballistic light transmission versus thickness for samples with lateral dimension of 15\,$\mu$m. a) The experimental data are presented with the black markers, while the red solid line shows the fitting to Lambert-Beer law with a scattering mean free path of $\ell\textsubscript{sc}=2.6\,\pm\,1.5\,\mu$m. The fitting error is 95\% confidence interval, shown with the black dashed line.
b) The color graph depicts ballistic light transmission for the structures as a function of the rods/volume (x-axis) and thickness (y-axis). The ballistic light attenuates with thickness, as expected. The largest decay of the ballistic light with thickness appears for 400 rods.}
\label{fig:lsc}
\end{center}
\end{figure}

\textbf{Figure \,\ref{fig:lsc}}(a) shows the physical thickness dependence of the ballistic light transmission in the series of structures with lateral dimensions of 15\,$\mu$m and 400 
inscribed rods. 
{\color{black}The ballistic transmission decreases exponentially with increasing thickness as $ I(\theta=0,L)=I\textsubscript{0}(\theta=0) \exp(-L/\ell\textsubscript{sc})$.}
We used non-linear least squares fitting of the experimental data 
and extract $\ell\textsubscript{sc}=2.6\,\pm1.5\,\mu$m where the error is the 95\% confidence interval. 
{\color{black}This value confirms that our media are in the multiple-scattering regime as $L \approx 5\times \ell\textsubscript{sc}$. 
}

In the color graph of Figure\,\ref{fig:lsc}(b) we summarize the ballistic light measurements for the series with lateral dimensions of 15\,$\mu$m.
On the x-axis, the thickness of the structure increases, while the y-axis shows the number of rods composing the full structures.
{\color{black}We observe the sharpest thickness-dependent drop in ballistic light transmission for structures composed of 400\,rods, which signifies the shortest $\ell\textsubscript{sc}$ among the fabricated structures.}

\begin{figure}[tbp!]
\includegraphics[width=0.9\linewidth]{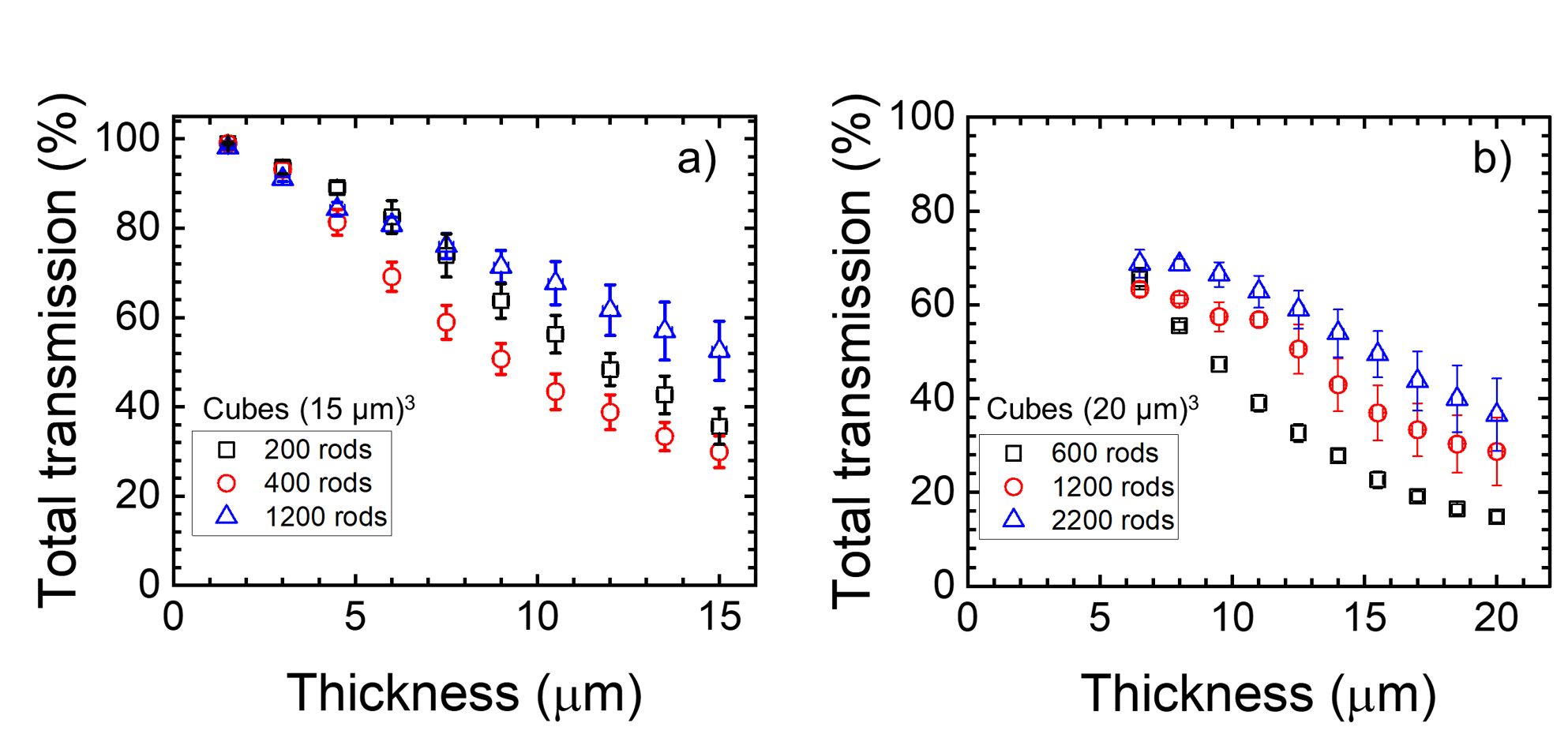}
\begin{center}
\caption{ Total transmission of the samples as a function of their thickness. In a) three series of samples are presented with different markers, corresponding to density of rods of 200, 400 and 1200\,rods and with lateral dimensions of 15\,$\mu$m. In b) three more cases presented, 600, 1200 and 2200\,rods and with lateral dimensions of 20\,$\mu$m. In both cases the media exhibit strong attenuation of the total transmission that can be tuned with the design parameters.
}
\label{fig:TT}
\end{center}
\end{figure}

\begin{figure}[tbp!]
\includegraphics[width=0.5\linewidth]{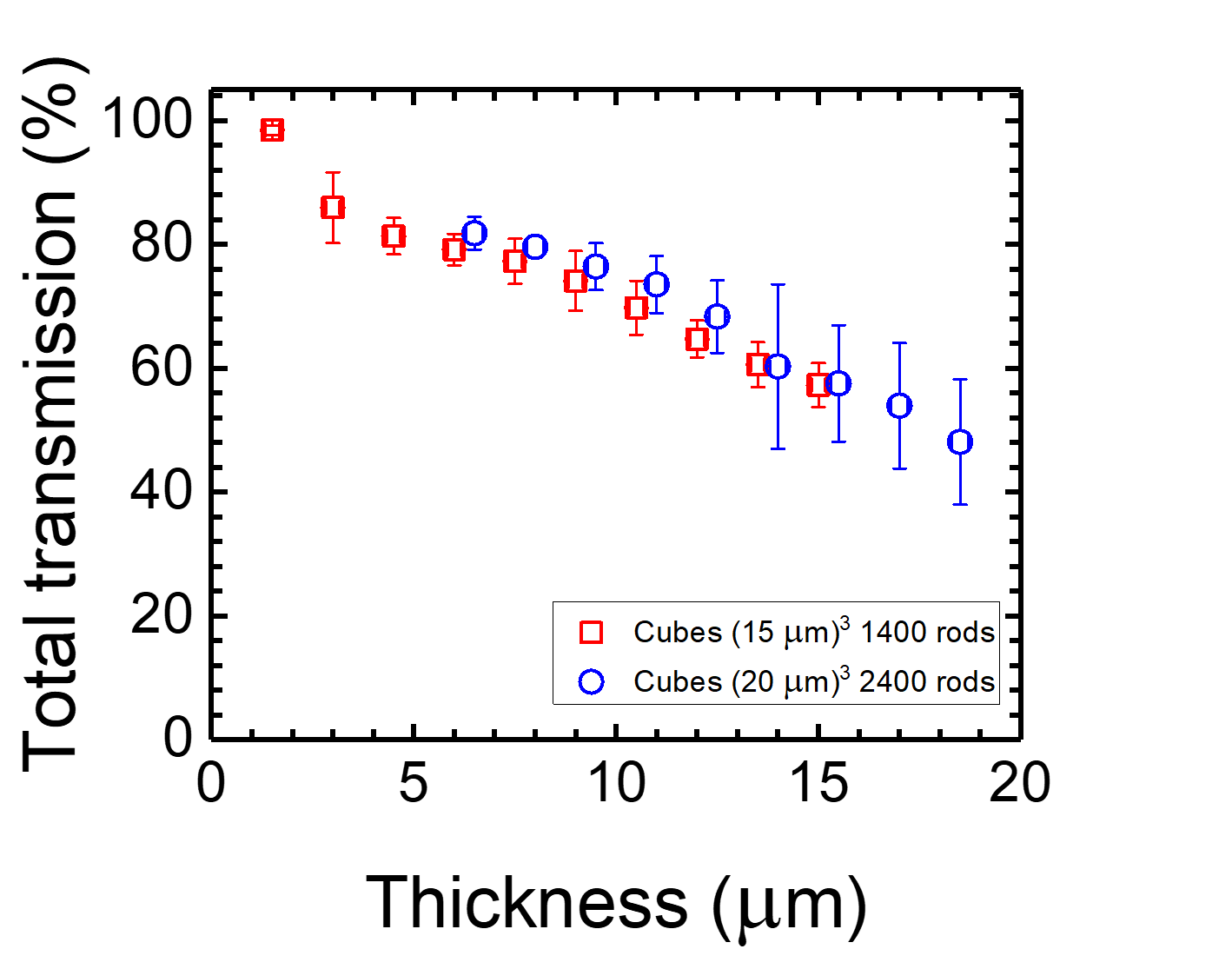}
\begin{center}
\caption{The total transmission values for structures of different density of rods over volume but the same filling fraction. The red markers correspond to experimental measurements series with 1500 rods in a $(15\,\mu{\rm m})^3$ cube while the blue markers correspond to series with 2400 rods in a $(20\,\mu{\rm m})^3$ cube. The filling fraction for the two series is 93\% and both trends match well within error bar, proving consistency in our fabrication and design. }
\label{fig:rho-same}
\end{center}
\end{figure}

\begin{figure}[tbp!]
\includegraphics[width=0.9\linewidth]{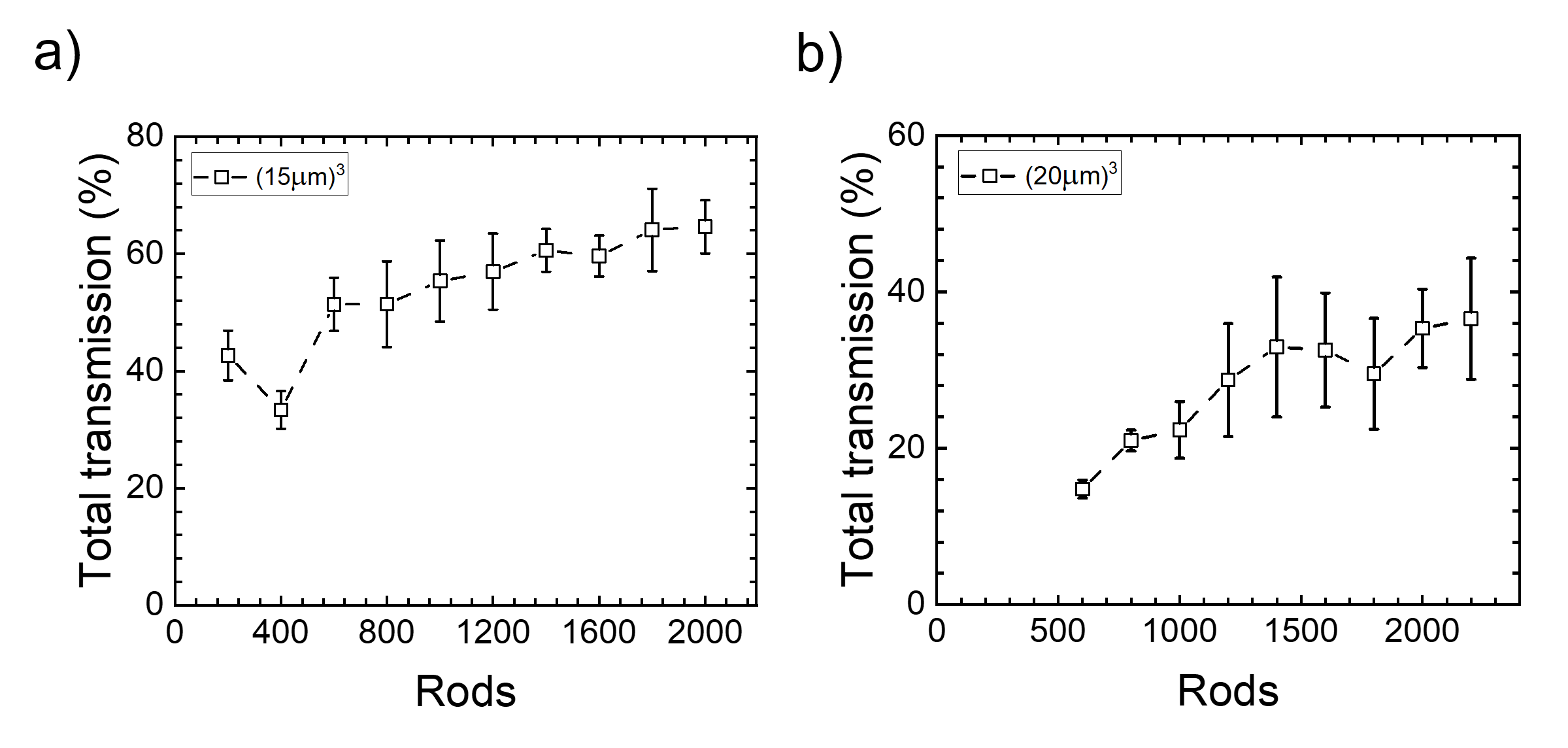}
\caption{Total transmission of the samples as a function of the density of rods over volume for given thickness. At the plot a), the markers denote the sample size for given volume (15\,$\mu$m)\textsuperscript{3} and increasing number of rods. At the plot b) the markers refer to the total transmission measurements versus increasing density of rods over volume for samples of volume (20\,$\mu$m)\textsuperscript{3}. Both graphs show agreement with figure \ref{fig:fill_neff}, where the lower transmission appears for filling fraction of 50\%. }
\label{fig:fill}
\end{figure}

The total transmission measurements are shown in \textbf{Figure\,\ref{fig:TT}} for structures with edge lengths of 15\,$\mu$m and 20\,$\mu$m in (a) and (b), respectively.
The total transmission decreases with increasing thickness of the structure. The two series show a different behavior for increasing thickness, which we attributed to the difference in the scattering mean free path and anisotropy values.
The transmission decrease is strongest for structures with 400 rods in (a) and 600 rods in (b).
Referring to Figure \,\ref{fig:fill_neff}, we remark that this is at $\sim$50\% filling fraction. This is an interesting contrast compared to samples made from spheroidal scatterers that find their maximum photonic strength at smaller values of filling fraction $\sim$30\%.
In \textbf{Figure\,\ref{fig:fill}}, we present the measured total transmission as a function of the filling fraction, for $15\,\mu$m and $20\,\mu$m lateral dimensions.
The lowest total transmission in our media appears at the 400 rods/(15\,$\mu$m\textsuperscript{3}) and 600 rods/(20\,$\mu$m\textsuperscript{3}), which corresponds to filling fractions of 54\,\% and 49\,\%, respectively. 
These values of the filling fraction indeed correspond approximately to maximizing the air-polymer interfaces while they are slightly higher than the filling fraction of $\approx 38\,\%$ that maximize photonic interaction for systems composed of spherical scatterers \cite{Busch1994, Vos2001}. 
To check the consistency of the transmission data for different linked dimensions at identical filling fractions ($\approx 93\%$), two total transmission series are made: for the 15\,$\mu$m cubes with 1400 rods 
and for the 20\,$\mu$m cubes with 2400 rods.
The lateral dimensions and the number of rods in these structures differ, but since the filling fractions coincide, the light transport is expected to be similar. 
This consistency is validated from \textbf{Figure\,\ref{fig:rho-same}} and highlights that the design method is effective for the chosen finite lateral dimensions and scatterer geometry.

%
%
%
%

The total transmission $T\textsubscript{z}$ for a sample of thickness $z$,
was measured by recording the total power of the light $I\textsubscript{z}$\ that was transmitted through the structure with given thickness $z$. 
The reference power $I\textsubscript{0}$ was measured on a bare glass substrate. 
The ratio of the total power for given thickness $z$ and the reference $I\textsubscript{0}$ is the total transmission of the sample,  $T\textsubscript{z}=I\textsubscript{z}/I\textsubscript{0}$.
{\color{black}To extract $\ell\textsubscript{sc}$, the ballistic light attenuation $B_{z}$ is measured as the ratio  $B\textsubscript{z}=I\textsubscript{z}(\theta = 0)/I\textsubscript{0}(\theta = 0)$, where $\theta$ is scattering angle with $\theta = 0$ along the propagation direction.} 


{\color{black}
The transport properties of a scattering medium can be derived using the radiative transfer equation, which can be solved analytically only for simple geometries.
Given the complexity of the nanofabricated structures that feature substrate reflections and finite size of the structures, it is impossible to exactly solve the radiate transfer equation.
Therefore, unambiguous estimation of $\ell\textsubscript{tr}$ or $g$ is technically demanding due to the lack of simple-to-use models that can be employed for fitting the measured transmission values.
In such a case, several numerical approaches can be used to extract the transport parameters from experimental data \cite{Hulst2012,Yang2004,Cheong1990}. 
Light transport Monte Carlo (MC) methods have been successfully employed to numerically solve complex geometries, finite boundary conditions, and even amplifying scattering media \cite{Prahl1989,Uppu2015}.
We employ MC simulations together with a least-square regression routine to fit the measured total transmission values and robustly extract $\ell\textsubscript{tr}$ as described below.

We perform MC simulations that numerically solve the radiate transfer equation \cite{Uppu2013,Meretska2018} for our sample geometry assuming an homogeneous distribution of anisotropic scatters in a sample volume as defined by our fabricated sample geometry.
Further, we also include Fresnel reflections from the sample substrate in the simulations.
Using MC simulations, we compute a look-up table of total transmission values over a wide parameter span of $\ell\textsubscript{sc}$ and $g$.
In our computations, the parameter span was set to a range of possible values of $\ell\textsubscript{sc}$ and $g$. For example, we use $\ell\textsubscript{sc} \in [0.1,300] \,\mu$m and $g \in [-1,1]$.
As the total transmission is a smooth function in this two-dimensional parameter space, the computed transmission values could be interpolated on a finer grid as required.
Such look-up tables were created for varying sample thicknesses at a given lateral size.
The MC estimates together with the measured total transmission data are then used to extract the best fit $\ell\textsubscript{sc}$ and $g$ using standard least-squares minimization routine.}
Note that the value derived here is an approximate value that is characteristic for transport in the $z$ direction averaged over the entire volume. Due to the elongated cross-section in the $z$ direction and the overall complexity of the rod network, the anisotropy requires a tensor description \cite{Cheong1990,Sapienza2004f}. Further study of the influence and control of anisotropy in DLW written structures is an intriguing direction for future research.

For a wavelength of 633\,nm the MC estimates for the 400 rods yield a scattering mean free path value $\ell\textsubscript{sc}=1.8\,\pm\,0.5\,\mu$m and $g=0.85\,\pm\,0.04$, corresponding to $\ell\textsubscript{tr} = 11.8\,\pm\,4.5\,\mu$m.
For the structure with 600 rods, MC estimates a scattering mean free path  $\ell\textsubscript{sc}=2.8\,\pm\,$0.3\,$\mu$m and $g=0.65\,\pm\,$0.04, which corresponds to $\ell\textsubscript{tr} = 8.3\,\pm\,1.2\,\mu$m.
We summarize the various results in \textbf{Table \ref{tab:values}}.
Comparing the results leads to a few observations: The 400 rods structure yields a smaller $\ell\textsubscript{sc}$ compared to the 600 rods structure. However, due to their respective value for the anisotropies, the corresponding order for their $\ell\textsubscript{tr}$ is inverted. 
We theorize that this occurs because at a higher density of rods we obtain a smaller value of the anisotropy. The more rods per volume, the less pronounced their geometrical features are, resulting in washing out the high anisotropy values. Vice versa this means that the scattering mean free path together with the anisotropy can be used to influence the transport mean free path. For us the most important conclusion is that the DLW structures are diffusive since their thickness surpasses the transport mean free path $L>\ell\textsubscript{tr}$, while possessing $\mu$m range $\ell\textsubscript{sc}$ and large $g$.
\\

\begin{table}[ht!]
 \caption {Transport parameters estimated by fitting the measured total transmission with the Monte Carlo simulations. Apart from the scattering mean free path $\ell\textsubscript{sc}$, the anisotropy $g$ and the transport mean free path $\ell\textsubscript{tr}$, also the optical thickness is listed.} \label{tab:values} 
 \begin{center}
 \begin{tabular}{@{}ccc@{}}
\hline
Parameter & 400 rods/$(15\,\mu$m$)^3$ & 600 rods/$(20\,\mu$m$)^3$ \\ 
\hline
$\ell_{sc}$ & 1.8\,$\pm$\,0.5\,$\mu$m & 2.8\,$\pm$\,0.3\,$\mu$m \\
$g$ & 0.85\,$\pm$\,0.04\, & 0.65\,$\pm$\,0.04\,\\
$\ell_{tr}$ & 12\,$\pm$\,4\,$\mu$m & 8\,$\pm$\,1\,$\mu$m \\ 
$L_{opt}$ & 1.2\,$\pm$\,0.3 & 2.5\,$\pm$\,0.1 \\ 
\hline 
\end{tabular}
\end{center}
\label{table}
\end{table}
%
%
%
%
%
%
%
%
%
%
%
\section{Conclusion}
We have implemented a DLW method to fabricate small deterministic optical multiple scattering media with optical thickness $\frac{L}{\ell\textsubscript{tr}} >1$. The fabrication process allows full control over the position and shape of the scatterers. 
We show that one can tune the density of scatterers and accordingly vary the transport mean free path. 
We demonstrate that our best design has a scattering mean free path of $\ell\textsubscript{sc}$=1.8\,$\pm\,0.5\,\mu$m. The deterministic nature of fabrication 
opens up the possibility for new fundamental studies of light propagation in scattering media.
This permits validation of various fundamental and applied aspects of light scattering for a given disordered structure, something that was not possible until now for optical frequencies \cite{Choi2011}.
In the future we want to investigate the reproducibility of the method to study the clonability of multiple scattering media as optical physical unclonable functions \cite{Goorden2014}.

\section{Experimental Section}
\subsection{Direct laser writing}

The structures were fabricated with a direct laser writing system \cite{Fischer2013} (Nanoscribe Professional GT, {\color{black}
using polymer photoresist Nanoscribe  IP-G with a refractive index of 1.51 for a wavelength of 633\,nm \cite{Gissibl2017}).}
The photoresist is a gel and its high viscosity ensures that the features in the structures do not drift during the writing processes, thereby minimizing 
deformation of the structures.
The illumination dose was set to be 14\% higher than the polymerization threshold (laser beam power 8.90\,mW and piezomotors scan speed of 140\,$\mu$m/s). 
This {\color{black} higher dose} results in rods of an average thickness of 500\,nm instead of 100\,nm that one could reach with a smaller illumination dose. 
Although the choice of a dose closer to the polymerization threshold would provide finer features \cite{Guney2016}, the slightly higher dose provides two advantages in the nanofabrication of those media. 
Firstly, thicker rods provide more robust mechanical support of the structure without any additional walls, thereby enabling self-supported structures unlike earlier demonstrations \cite{Deubel2004}.
The second advantage of a higher dose is the reduction of fabrication disorder through the complete cross-linking of polymer chains across the entire volume 
\cite{Lee2008}. 

\subsection{Optical setup}	

\begin{figure}[htbp]
\begin{center}
\includegraphics[width=0.6\linewidth]{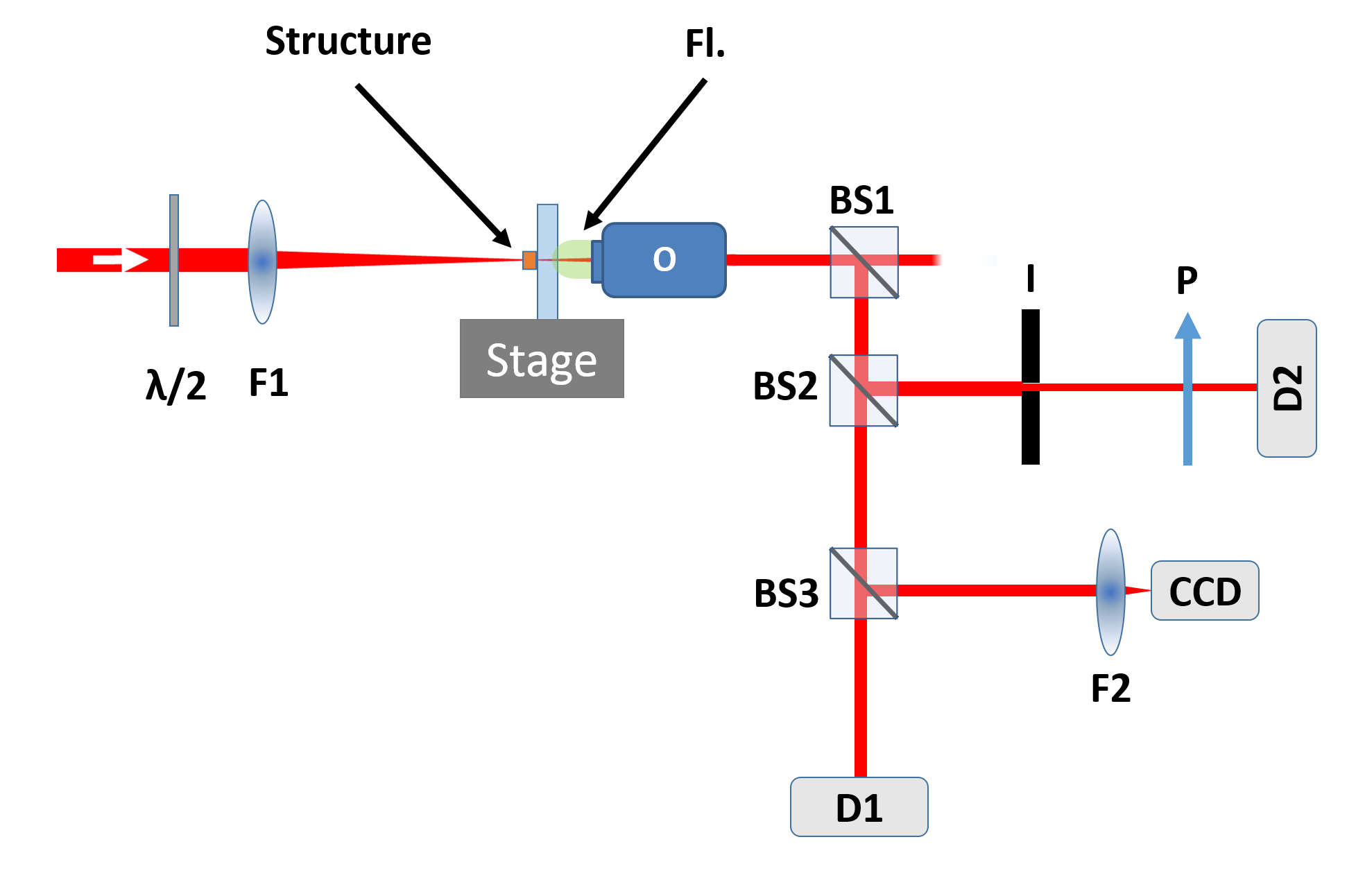}
\caption{Schematic of the set up for measuring the total transmission and ballistic light attenuation.  a) A continuous wave He:Ne laser at 633\,nm is expanded to a diameter of 3.1 cm and focused on a single structure with an aspheric lens F1 with a focal length of 100\,mm. A microscope oil immersion (Fl.) objective with NA=1.4 collects the transmitted light from the glass surface.
Illumination of a specific structure is controlled with the use of a 3-axis piezo positioning stage. We use a CCD camera to confirm correct positioning while the total transmission is recorded with a power meter (D1). 
The light that is reflected on BS2 is used to evaluate the ballistic light attenuation. An iris with a polarizer aligned parallel to the polarization of the laser beam filters out the scattered light. The filtered light intensity is measured with a power meter (D2).}
\label{fig:setup}
\end{center}
\end{figure}

The optical scattering strength in our samples is quantified by extracting the transport mean free path $\ell\textsubscript{tr}$\ using total transmission measurements \cite{Garcia1992}. 
The experimental set-up is shown in \textbf{Figure\,\ref{fig:setup}}. 
The laser beam from a linearly-polarised He-Ne laser ($\lambda=633\,$nm, beam diameter of $3\,$mm) is expanded $10\times$ with a beam expander and focused with a plano-convex lens (focal length = +100\,mm) onto the sample. 
The focused beam has a full-width at half-maximum of 4 $\mu$m, smaller than the lateral dimensions of 15 and 20\,$\mu $m of our structures. 
An oil-immersion objective(numerical aperture NA=1.4) is used to collect the light transmitted light through the photonic scattering medium.
We position each structure with a 3-axis piezo positioning stage, while viewing with a CCD camera.
The light collected by the objective is split with three non-polarizing 50:50 beam splitters (BS1--3). 
The transmitted part of BS3 is sent to power meter(D1
) to obtain the total transmission. 
The reflected part of BS2 is used to monitor the ballistic light. 
For this reason it is filtered with an adjustable iris.
A polarizer filters out multiple-scattered light, allowing only polarization components parallel to the original beam polarization to be transmitted to the power meter (D2
).

\medskip
\textbf{Acknowledgments}
We thank Cock Harteveld, Ad Lagendijk, and Matthijs Velsink for
discussions and support. This work is financially supported by the
Nederlandse Wetenschaps Organisatie (NWO) via QuantERA QUOMPLEX (Grant No. 68091037), Vici (Grant No. 68047614) and NWA (Grant No. 40017607), by STW project 11985, and by the FOM program 'Stirring of light'. 
%
%
%
\medskip

\end{document}